\documentclass[aps,twocolumn,prd,floats,nofootinbib,showpacs]{revtex4}

\usepackage{graphicx}
\usepackage{amsmath}
\usepackage{amsfonts}
\usepackage{amssymb}
\def\mhsm{m_{h_{SM}}}
\def\hpm{H^\pm}
\def\hp{H^+}
\def\hm{H^-}
\def\mhpm{m_{\hpm}}
\def\wmp{W^{\mp}}

\def\pbi{~\mbox{pb}^{-1}}

\def\ma{m_A}
\def\mh{m_h}
\def\mhh{m_H}

\def\vev#1{\langle #1 \rangle}

\def\beq{\begin{equation}}
\def\eeq{\end{equation}}
\def\bea{\begin{eqnarray}}
\def\eea{\end{eqnarray}}
\def\ie{{\it i.e.}}

\def\lsim{\mathrel{\raise.3ex\hbox{$<$\kern-.75em\lower1ex\hbox{$\sim$}}}}
\def\gsim{\mathrel{\raise.3ex\hbox{$>$\kern-.75em\lower1ex\hbox{$\sim$}}}}
\def\ifmath#1{\relax\ifmmode #1\else $#1$\fi}

\def\fbi{~{\mbox{fb}^{-1}}}

\def\pb{~{\mbox{pb}}}
\def\br{BR}
\def\gev{~{\mbox{GeV}}}

\def\to{\rightarrow}

\def\hsm{h_{SM}}

\def\lam{\lambda}
\def\sig{\sigma}
\def\anti{\overline}

    \def\fillboxx#1#2{\hbox to #1{\vbox to #2{\vfil}\hfil}   }

\def\gev{~{\rm GeV}}
\def\gam{\gamma}
\def\tanb{\tan\beta}
\def\cotb{\cot\beta}

\def\Eq#1{Eq.~(\ref{#1})}
\def\cosb{\cos\beta}
\def\sinb{\sin\beta}
\def\cosa{\cos\alpha}
\def\sina{\sin\alpha}

\def\anti{\overline}

\def\sigwjj{\sigma_{Wjj}}

\def\wp{W^+}
\def\wm{W^-}

\def\rgamgam{R_{\gam\gam}}
\def\rgamgama{\rgamgam^A}

\newcommand{ \slashchar }[1]{\setbox0=\hbox{$#1$}   
   \dimen0=\wd0                                     
   \setbox1=\hbox{/} \dimen1=\wd1                   
   \ifdim\dimen0>\dimen1                            
      \rlap{\hbox to \dimen0{\hfil/\hfil}}          
      #1                                            
   \else                                            
      \rlap{\hbox to \dimen1{\hfil$#1$\hfil}}       
      /                                             
   \fi}     
\begin{document}
\title{\boldmath A two-Higgs-doublet interpretation of a small Tevatron $Wjj$ excess
}

\author{John  F. Gunion}

\affiliation{ Department of Physics, University of California, Davis,
  CA 95616, USA}

\begin{abstract}
  We show that a $Wjj$ excess in Tevatron data could be explained in the
  context of the standard  non-supersymmetric two-Higgs-doublet model (2HDM) for
  appropriately chosen parameters.  Correlated signals in the
  $\gam\gam$ and $\wp\wm b\anti b$ final states are predicted and are
  on the verge of being detectable. The proposed model is most
  attractive if the cross section for the $Wjj$ excess is $\lsim 1-2\pb$.

\end{abstract}


\maketitle

The discrepancy between the CDF~\cite{Aaltonen:2011mk} and
D0~\cite{Abazov:2011rf} results implies considerable uncertainty as to
whether there is an excess of $Wjj$ events in the $M_{jj}\sim 140\gev$
region. Nonetheless, it is interesting to explore the different
theoretical approaches that could produce such an excess. Many
possibilities have appeared in the literature, including several Higgs
sector approaches. A probably incomplete summary is the following:
approaches based on $SU(2)$ doublet scalars with or without extra
singlets \cite{Wang, Babu, Dutta:2011kg,Segre:2011tn,Chen}; $Z$-prime
models \cite{Hooper, Cheung}; new colored state models
\cite{XPWang,Dobrescu,Carpenter:2011yj}; supersymmetry models
\cite{Kilic, Sato}; technicolor models \cite{Lane}; string theory
models \cite{stringy}; and within the context of the Standard model
\cite{He, Sullivan:2011hu,Plehn:2011nx}. This Letter demonstrates that
the excess could be explained by the simplest non-supersymmetric
two-Higgs-doublet (2HDM) model {\it with completely standard Yukawa
  coupling structure}.  The model predicts correlated signals in the
$\gam\gam$ and $\wp\wm b\anti b$ final states that are on the verge of
detection.

We begin with a general overview of the approach. We employ a
two-Higgs-doublet model of Type-II (a convenient summary appears in
the HHG~\cite{hhg}).  In the context of the 2HDM, the masses of the
light and heavy CP-even Higgs bosons, $h$ and $H$, of the CP-odd Higgs
boson, $A$, and of the charged Higgs boson $H^\pm$, as well as the
value of $\tanb=v_u/v_d$ ($v_{u,d}=\vev{H^0_{u,d}}$ where $H^0_{u,d}$
couple to up-type, down-type quarks, respectively) and the CP-even
Higgs sector mixing angle $\alpha$ can all be taken as independent
parameters, whose values will determine the $\lam_i$ of the general
2HDM Higgs potential.

To obtain a $Wjj$ signal with Tevatron cross section of order $\gsim
1\pb$, the first ingredient is to note that the cross section for
$gg\to A$ is highly enhanced at a given $\ma$ relative to the cross
section for a SM Higgs boson at $\mhsm=\ma$ when $\tanb< 1$. The $Wjj$
signal derives from the (dominant) $A\to \hpm\wmp$ decay channel with
$\hpm\to cs$. Note that this particular mode does not contain $b$
quarks, as consistent with the CDF observations.\footnote{However,
  $\hpm \to t^* b$ has a large branching fraction, as discussed later,
  but since $t^*\to W b$, this channel will not lead to a $jj$
  resonance signal.} Using the predicted value of $BR(\hp\to cs)\sim
0.2$ for $\mhpm\sim 140\gev$ when $\tanb$ is small, one finds that a
cross section for $gg\to A\to \hpm \wmp\to cs\wmp$ as large as the CDF
value of $\sim 4\pb$ can only be achieved for $\ma\in[250,300]\gev$ if
$\tanb\lsim 1/10$, a domain for which the top-quark Yukawa coupling is
non-perturbative, $\alpha_t\equiv \lam_t^2/(4\pi)>1$.  However, a
smaller $Wjj$ cross section of order $1-2\pb$ is possible for
$\alpha_t\sim 1$. We now present more details.

In the 2HDM there are only two
possible models for the fermion couplings that naturally avoid flavor
changing neutral currents (FCNC), Model I and Model II. In Model II,
our focus here, the tree-level couplings of the Higgs bosons are:
\vspace*{-.12in}
\beq
\begin{array} {|c|c|c|c|}
\hline
\ & h & H & A \cr
\hline 
t\anti t & {\cosa\over\sinb} & {\sina\over\sinb} & -i\gam_5 \cotb  \cr
\hline
b\anti b & -{\sina\over\cosb} & {\cosa\over\cosb} & -i\gam_5\tanb \cr
\hline
WW,ZZ & \sin(\beta-\alpha) & \cos(\beta-\alpha) & 0 \cr
\hline
\end{array}
\label{coups}
\vspace*{-.12in}
\eeq 

We will fix $\alpha$ relative to $\beta$ by requiring that the $h$ be
SM-like, \ie\ $\sin(\beta-\alpha)=1$. We also choose $\mh=115\gev$ for easy
consistency with precision electroweak data. If the $\lam_i$ of the
Higgs potential are kept highly perturbative, the decoupling limit, in
which $\mhh,\mhpm\to\ma$ and $\sin^2(\beta-\alpha)\to 1$, sets in
fairly quickly as $m_A$ increases~\cite{Gunion:2002zf}. To describe a
$Wjj$ excess requires that $\mhpm<\ma$ (but $\mhh\sim\ma$ is useful to
enhance the signal), implying that the decoupling limit does not apply
at the masses of interest.  This requires that several of the $\lam_i$
are substantial but still below the $\lam_i^2/(4\pi)\sim 1$ beginning
of the non-perturbative domain.

\begin{figure}
\includegraphics[height=0.5\textwidth,angle=90]{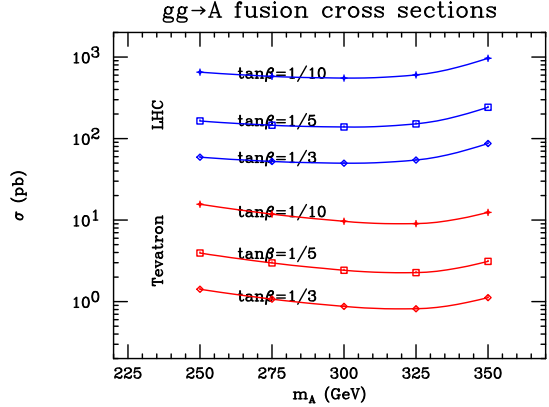}
\vspace*{-.5in}\caption{Tevatron and LHC cross sections for $gg\to A$
  for representative $\tanb<1$ values.}
\label{ggafusion}
\end{figure}

Looking at \Eq{coups}, it is apparent that the cross section for
$gg\to A$ can be large when $\cotb>1$.  It is also useful to recall
that the fermionic loop function for the $A$ is substantially larger
than that for the $H$ (the CP-even Higgs that could contribute to the
$Wjj$ excess if the $h$ is SM-like); {\it e.g.}  asymptotically
$F_{1/2}^A(\tau)\to 2$ in comparison to $F_{1/2}^{H}(\tau)\to -4/3$
when $\tau=4m_f^2/m_A^2\to \infty$, implying a cross section gain by a
factor of $9/4$ for $A$ vs. the $H$ in the heavy fermion mass
limit. We have computed the $gg\to A$ (and $gg\to H$) cross section using
HIGLU~\cite{Spira:1996if} and a private program and obtained
essentially the same results. Results for $\sigma(gg\to A)$ are
plotted in Fig.~\ref{ggafusion}. These results include NLO and NNLO
corrections as in HIGLU. Some useful benchmark numbers for
$\ma=250\gev$ are 
\beq
\begin{array} {|l|l|l|l|}
\hline
\tanb & 1/3 & 1/5 & 1/10 \cr
\hline
\sig(gg\to A)_{Tevatron} & 1.4\pb & 3.9\pb & 15.7\pb \cr
\hline
\sig(gg\to A)_{LHC} & 59.1\pb & 164.3 \pb & 652.9 \pb \cr
\hline
\end{array}
\label{ggatab}
\eeq

\begin{figure}
\vspace*{-.25in}
\includegraphics[height=0.5\textwidth,angle=90]{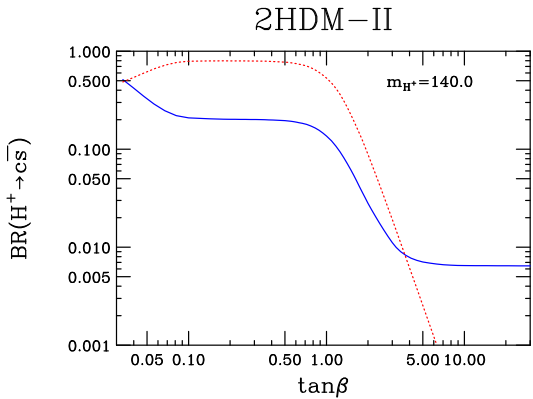}
\vspace*{-.5in}\caption{$\br(\hp\to c\anti s)$ (solid blue) and
  $\br(\hp\to t\anti b)$ (red dots) as a function of $\tanb$ for
  $\mhpm=140\gev$ and Model~II couplings. Inclusion of off-shell decay
  configurations is essential for the $t\anti b$ final state.}
\label{brhptocs}
\end{figure}

\begin{figure}
\includegraphics[height=0.5\textwidth,angle=90]{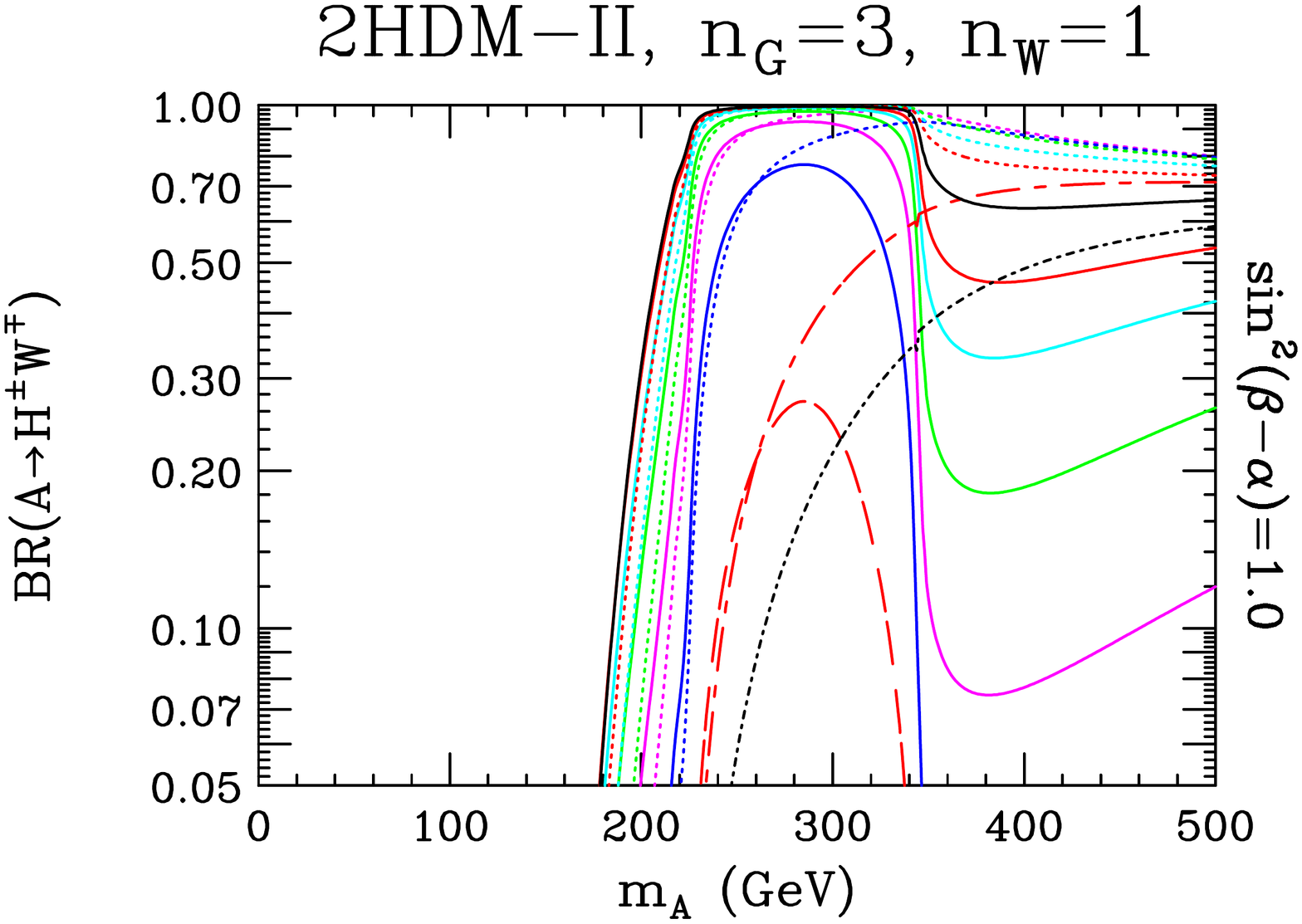}
\vspace*{-.5in}\caption{$\br(A\to \hpm\wmp)$ as a function of
 $\ma$ for $\mhpm=140\gev$ and Model~II couplings. In this and
 subsequent plot for the $A$, we have taken $m_H=140\gev$.
The legend is as follows: solid
  black$\to\tanb=1$; red dots$\to\tanb=1.5$; solid
  red$\to\tanb=1/1.5$; cyan dots$\to\tanb=2$; solid
  cyan$\to\tanb=1/2$; green dots$\to\tanb=3$; solid
  green$\to\tanb=1/3$; magenta dots$\to\tanb=5$; solid
  magenta$\to\tanb=1/5$; blue dots$\to\tanb=10$; solid
  blue$\to\tanb=1/10$; long red dashes plus dots$\to\tanb=30$; pure
  long red dashes$\to\tanb=1/30$; black dotdash$\to\tanb=50$. This and
  subsequent 
  figures must be viewed in color in order to resolve the different
  $\tanb$ cases.  Results plotted include off-shell
  decay configurations. $n_G=3$, $n_W=1$ means 3 generations, no
  sequential $W'$.}
\label{bratohpmwmp}
\end{figure}

We define the effective $Wjj$ cross section for a Higgs boson $X$:
\beq
\sigwjj^X\equiv \br(X\to\hpm\wmp) \br(\hp\to
c\anti s)\sig(gg\to X),
\eeq
where $X=A$ and $X=H$ are the relevant Higgs bosons.  As a benchmark
to keep in mind, we will suppose that $\sigwjj^A\sim 1\pb$ is
appropriate for describing the Tevatron $Wjj$ excess.  $\br(\hp\to
c\anti s)$ (computed privately and using HDECAY~\cite{Djouadi:1997yw})
is displayed in Fig.~\ref{brhptocs} where we see that a value of $\sim
0.22$ applies for $\tanb\in[1/10,1/3]$.  For $\ma=250\gev$,
Fig.~\ref{bratohpmwmp} shows that $\br(A\to \hpm\wmp)\sim
0.95,0.874,0.64$ for $\tanb=1/3,1/5,1/10$ (the solid green, magenta,
blue lines), respectively.  For $\ma=250\gev$ we then obtain $\br(A\to
\hpm\wmp)\br(\hp\to c\anti s)\sim 0.21,0.19,0.14$ for
$\tanb=1/3,1/5,1/10$.  Using the $\sig(gg\to A)$ cross sections of
\Eq{ggatab}, for $\ma=250\gev$ we find $\sigwjj^A(Tevatron)\sim
0.3\pb,\ 0.75\pb,\ 2.2\pb$ for $\tanb=1/3,\ 1/5,\ 1/10$,
respectively. The corresponding values of $\alpha_t$ are $0.63,\
1.75,\ 7$. Only the latter is uncomfortably (but not drastically)
non-perturbative, implying a preference for $\sigwjj^A\lsim 1\pb$. It
is quite important to note that the main reason that $\sigwjj^A$ is
not larger is the small value of $\br(\hp\to c\anti s)$ that is a
consequence of the dominance of {\it off-shell} $\hp\to t^*\anti b$
decays for $\mhpm = 140\gev$. (This dominance decreases rapidly if
$\mhpm$ is decreased; for $\mhpm$ significantly lower that $140\gev$
higher $\sigwjj^A$ would thus be achieved.)  For $\ma\gsim 300\gev$,
$\sigwjj^A$ is about 50\% smaller than the $\ma=250\gev$ values quoted
above, see Fig.~\ref{ggafusion}.

\begin{figure}
\includegraphics[height=0.5\textwidth,angle=90]{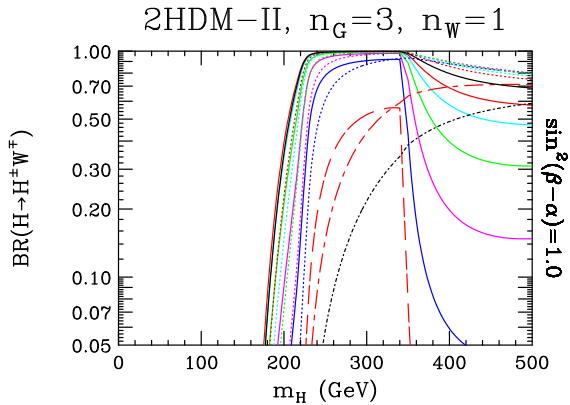}
\vspace*{-.5in}\caption{$\br(H\to \hpm\wmp)$ as a function of
 $\ma$ for $\mhpm=140\gev$ and Model~II couplings. In this and
 subsequent plots for the $H$, we have taken $\ma=200\gev$. The legend is as in
Fig.~\ref{bratohpmwmp}. }
\label{brHtohpmwmp}
\end{figure}

As apparent from \Eq{ggatab}, $\sig(gg\to A)$ is much larger at the
LHC.  Focusing on $\ma=250\gev$ and including the earlier quoted
$\br(A\to\hpm\wmp)\br(H^+\to c\anti s)$ values of $0.21,\ 0.19,\ 0.14$
we obtain $\sigwjj^A(LHC)=12.4\pb,\ 31.2\pb,\ 91.4\pb$ for
$\tanb=1/3,\ 1/5,\ 1/10$, respectively.  The number of $Wjj$ events
will be enormous for the soon-to-be-achieved $L=1\fbi$. We anxiously
await the appropriate LHC analyzes.

It is, of course, interesting to assess the extent to which $gg\to
H\to \hpm\wmp$ with $\hp\to c\anti s$, $\hm\to \anti c s$ could
contribute to the $Wjj$ final state (recall that the $h$ is taken to
be light and SM-like so that only the $H$ is relevant for the $Wjj$
excess). We have already noted that $\sigwjj^H<\sigwjj^A$ due to the
smaller fermionic loop function.  Actual ratios at the Tevatron are:
$\sigwjj^A/\sigwjj^H\sim 2.6,\ 3.0,\ 5.0$ for $\ma=\mhh=250,\ 300,\
350\gev$. Meanwhile, the $\br(H\to \hpm\wmp)\br(\hp\to c\anti s)$
values are similar to those quoted for the $A$.  Thus, for the
preferred $\mhh\in[250-300]\gev$ mass range, the $H$ would yield a
$Wjj$ signal of order $30\%-40\%$ of the $A$ result.  If the $H$ and
$A$ are not fairly degenerate, this would yield a somewhat spread out
net $Wjj$ signal, despite the $\lsim 1\gev$ total widths of the $A$
and $H$ (for the $\tanb$ values being discussed), given the
experimental $M_{jj}$ resolution of order $15\gev$. This is perhaps
suggested by the absence of any distinct peaking in the $Wjj$ mass in
the data.  Another interesting point is that in this model with $\mhh$
not very different from $\ma$, there would be no signal in the $Zjj$
channel due to the absence of $H\to AZ$ and $A\to HZ$ decays.

\begin{figure}
\includegraphics[height=0.5\textwidth,angle=90]{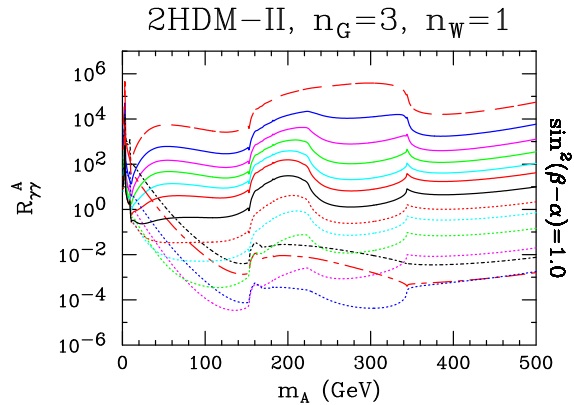}
\vspace*{-.5in}\caption{$\rgamgam$ for the 2HDM-II $A$. The legend is as in
Fig.~\ref{bratohpmwmp}. This figure takes account of all the $A$ decay
modes, including especially $A\to\hpm\wmp$ as well as $A\to t\anti t$
(off-shell) decays.}
\label{hAII}
\end{figure}

\begin{figure}
\includegraphics[height=0.5\textwidth,angle=90]{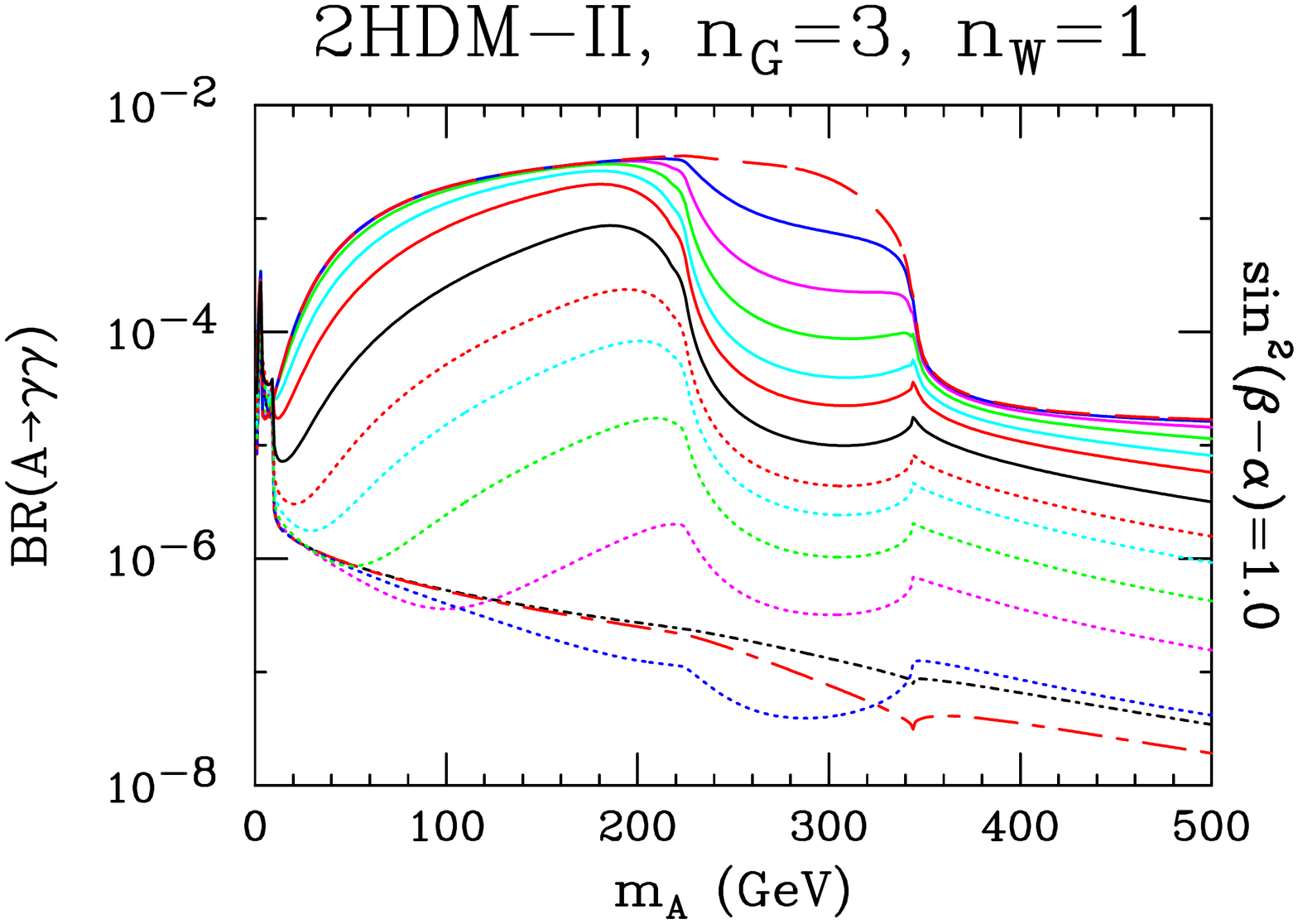}
\vspace*{-.5in}\caption{$\br(A\to \gam\gam)$ for the 2HDM-II $A$ after
  including $A\to \hpm\wmp$ and $A\to t\anti t$ off-shell decays in the present scenario. The legend is as in
Fig.~\ref{bratohpmwmp}. }
\label{braiigamgam}
\end{figure}

Other signals should be seen if the model is correct.  In particular,
as pointed out in~\cite{Gunion:2011ww}, there is a very large $A\to
\gam\gam$ signal for small $\tanb$.  A plot of $\rgamgama\equiv
[\Gamma_{gg}^A\br(A\to\gam\gam)]/[\Gamma_{gg}^{\hsm}\br(\hsm\to\gam\gam)]$
is given in Fig.~\ref{hAII}. For $\tanb\sim 1/3,1/5$ and $\ma\sim
250\gev$, $\rgamgama\sim 10^2,10^3$, respectively!  Such a signal will
soon be observed at the LHC if present and might also be observable
with current Tevatron data. To assess actual event rates one can
combine the actual branching ratio for $A\to \gam\gam$, plotted in
Fig.~\ref{braiigamgam} with the cross sections for $gg\to A$ plotted
in Fig.~\ref{ggafusion}.  

For example, for $\tanb=1/5$ and $\ma=250\gev$, in the case of the
Tevatron one finds $\sigma(gg\to A)\br(A\to \gam\gam)\sim 3.9\pb
\times 4.8\cdot 10^{-4}\simeq 1.9\times 10^{-3}\pb$, yielding $\sim 10$
events for $L=5.4\fbi$.  This must be compared to the number of events
expected in the SM. Ref.~\cite{Aaltonen:2010cf} performs an analysis
for $L=5.4\fbi$. Their net efficiency times acceptance is $\sim 0.12$,
implying a predicted number of $A\to\gam\gam$ events of order $1.2$.
The actual number of observed events is consistent with the SM
prediction, as shown in their Fig.~2. They set a
95\% CL limit of $\sigma\br(\gam\gam)\lsim 0.05\pb$ at
$M_{\gam\gam}=250\gev$, a factor of $\sim 25$ above our typical
prediction. At the LHC, the corresponding calculation is $\sigma(gg\to
A)\br(A\to\gam\gam)\sim 164\pb \times 4.8\cdot 10^{-4}=0.08\pb$.
For $L=36\pbi,1\fbi$ this yields $\sim 3,80$ events,
respectively. Ref.~\cite{cmsgravitonstudy} uses $L=36\pbi$ data 
to set a limit of $\sigma\times \br(\gam\gam)\lsim 0.7\pb$ at
$M_{\gam\gam}=250\gev$, a factor of about $8$ above the prediction for
the present scenario.  This shows that the present scenario for
obtaining a $Wjj$ excess will be strongly tested once the currently available
LHC data sets with $L=1\fbi$ are analyzed.

Of course, the $H$ also yields a large $\gam\gam$ signal (again of
order $30\%-40\%$ that of the $A$) that most probably would be
detected as a separate peak if $\mhh$ differs from $\ma$ by more than
$10\gev$, given the excellent $\sim 2\gev$ mass resolution in
$M_{\gam\gam}$ for the LHC detectors and given that the total $A$ and
$H$ widths are of order $1\gev$.

Finally, there is an interesting signal in the $gg\to A\to \hpm
\wmp\to t^*\anti b \wm+\anti t^* b \wp$ final state
deriving from the $\br(\hp\to t^* \anti b)\sim 0.7$ (off-shell) decays, see
Fig.~\ref{brhptocs}, where $t^*\to \wp b$. The resulting final states
of $\wp\wm b\anti b$ will not peak in either $Wb$ mass combination.
The cross section for this final state is, however, significant: for
$\ma=250\gev$ and $\tanb=1/5$, one finds $\sigma(WWbb)\sim 2.8 \pb$
compared to $\sigwjj^A\sim 0.75\pb$ and $\sigwjj^H\sim 0.28\pb$.
Although this $\sigma(WWbb)$ is somewhat smaller than that for direct
$t\anti t\to \wp\wm b\anti b$ production, it is still sizable and
might lead to some ``anomalies'' in the $\wp\wm b\anti b$ final state.
It would be very interesting to determine whether or not such
anomalies in the $\wp\wm b\anti b$ final state would have been noticed
in current data and, if not, how much LHC integrated luminosity would
be needed to detect them.  One should note that for this model to
achieve the CDF $Wjj$ cross section of $\sim 4\pb$ would imply an
anomalous $\wp\wm b\anti b$ final state cross section that is larger than
that coming directly from $t\anti t\to \wp\wm b\anti b$ production.

If a 4th generation is present with $m_{t',b'}\sim 400\gev$, then
$\Gamma_{gg}^A$ and, therefore, $\sigwjj^A$ is increased substantially
at any fixed $\tanb$. However, $\tanb$ is restricted to lie in the range
$\tanb\in[1/2,2]$ in order to keep $\lam_{t',b'}^2/(4\pi)\lsim 1$.  The
resulting $gg\to A$ rate is then more or less the same as for
$\tanb\in[ 1/3,1/5]$ with no 4th generation.

Enhanced $gg\to A$ cross sections also arise in a Model~I 2HDM if
$\tanb<1$.  However, the enhancement is not quite as great as for
Model~II. In addition, $\br(\hp\to c\anti s)\sim 0.13$ for $\tanb\in[
1/3,1/10]$.  As a result, the $Wjj$ cross section that can be achieved
in Model~I is smaller by about a factor of three as compared to that
achieved for the $Wjj$ final state in the case of Model~II.

To summarize, we have shown that if $\tanb$ is small then a Model~II
two-Higgs-doublet sector with $\ma$, and possibly $\mhh$, of order
$250\gev-300\gev$ can lead to a very interesting signal in the $Wjj$
final state that could match that seen by CDF at the Tevatron.  To get a
cross section as large as that originally claimed by CDF
would force one to $\tanb\lsim 1/10$, values for which the top-quark
Yukawa coupling is quite large and moderately
non-perturbative. However, a $Wjj$ signal with cross section of order
$1\pb$, as possibly consistent with a combination of CDF and D0 data,
is quite possible without entering into the domain of non-perturbative
top-quark Yukawas. Correlated signals in the $\wp\wm b\anti b$ and
$\gam\gam$ final states are expected.  These final states are
interesting targets for exploration in their own right. 
The  predicted correlations between the $Wjj$, $\wp\wm b\anti b$ and
$\gam\gam$ signals makes the model proposed
herein highly testable and points out the importance of taking into
account the latter types of signals in order to fully assess the
consistency of the model.  At the LHC, the predicted $Wjj$ cross
sections and those for the correlated signals are of order 40 times as
large as at the Tevatron.  As the integrated LHC luminosity approaches
$L=1\fbi$ the model will most probably be definitively eliminated or
confirmed.
As a final note, the masses for the $\mhpm$, $\ma$
and $\mhh$ needed to explain the possible $Wjj$ excess using the
approach described here cannot be achieved within the minimal
supersymmetric model context.

\vspace*{-.2in}
\acknowledgments 

JFG is supported by U.S. DOE grant No. DE-FG03-91ER40674. Thanks to
S. Chang for a critical examination of the paper and helpful comments.

\end{document}